\documentclass[twocolumn,showpacs,nofootinbib,aps,eqsecnum,prd,notitlepage,showkeys,10pt]{revtex4-1}

\usepackage{amssymb}
\usepackage{ulem}
\usepackage{amsmath}
\usepackage{graphicx}
\usepackage[usenames]{xcolor}
\usepackage{dcolumn}
\usepackage{hyperref}
\usepackage{amsfonts,epsfig,xspace}
\usepackage{pstricks,pst-node}
\usepackage{breqn}

\begin{document}

%\title{Effective Equations of the Quantum FRW Flat Universe in the Radiation Dominated Era}
\title{Effective FRW Radiation Dominated Era}
\author{Aureliano Skirzewski}
\author{Jes\'us Rodr\'iguez}
\affiliation{Centro de F\'{\i}sica Fundamental. Universidad de Los Andes. 5101 M\'erida. Venezuela.}

\begin{abstract}

We compute effective equations of the quantum FRW flat universe in the radiation dominated era at order $\hbar$, described in terms of Ashtekar variables employing a new method for the geometrical formulation of quantum mechanics. Additional terms of quantum nature  correct the classical equations of motion. As a consequence, the initial singularity of the classical model is removed and a Big Bouncing scenario takes its place. We also obtain an expression for the effective action of the model in terms of higher curvature invariants, leading us to  corrected Einstein equations for more general contexts.     

\end{abstract}

\maketitle

\section{Introduction}

Friedmann-Robertson-Walker cosmological models provide a successful description of the universe at large scale. In spite of that, it has technical and conceptual difficulties when the universe becomes smaller backwards in time. Yet, at small scales we expect quantum effects to  produce important modification to FRW's models.

Of course, quantum description  requires  a canonical formulation of the model as starting point.  
Here, we use Ashtekar's formulation of General Relativity \cite{Ashtekar}  for the kinematics of the cosmological model, with usual cosmological symmetries such as homogeneity, isotropy and spacial flatness \cite{reviewbojowald}.

A complete quantum treatment that includes information of the entire wave functions in Quantum Cosmology is a notoriously difficult problem, not only at the computational level but because of the interpretation and extraction of conclusions. 
On the other hand, effective theories allow us to describe some properties of the whole quantum theory by interactions of a finite number of degrees of freedom. 
Effective equations provide a simpler approach in terms of classical quantities and we will study them through the determination of the quantum correlations associated to central moments of a state \cite{PaperTesisAureliano}. More specifically, we will use techniques associated to classical mechanics in order to study quantum corrections to the cosmological model in the radiation dominated era.

To describe the behavior of the universe close to the Big Bang following FRW cosmology, we are going to study how quantum effects modifies physics in the radiation dominated era of the evolution of the universe. For this purpose, we start studying the Hamiltonian model proposed in \cite{Bojowald1}. The problem of time \cite{isham} is treated in the same way as \cite{Bojowald1} did. Then, we obtain the effective equations of the model, where quantum correction terms arise and study its solutions. Finally we analyze the physical implications of the results and discover that the big bang scenario is replaced by a big bounce. 

\section{Canonical Description of the Radiation Dominated Era}
\label{II}

\subsection{The Hamiltonian Constraint}

In a FRW scenario, if we assume homogeneity, isotropy and spatial flatness, the metric in a particular reference frame can be assumed to be of the form 
$$g_{\mu\nu}=-\delta^t_\mu\delta_\nu^t +a^2(t)\delta^a_\mu\delta^b_\nu,$$
where $a(t)$ is the cosmic scale factor that evolves in time according to Einstein's theory of gravitation.
The same model can be reformulated in Ashtekar variables, that are related to the cosmic scale factor and its time derivative, the extrinsic curvature \cite{reviewbojowald} $$A^{a}_{i}=c\delta^{a}_{i}$$ $$E^{i}_{a}=p\delta^{i}_{a}$$ with canonical gravitational variables $c=\gamma\dot{a}$ and $p=a^{2}$ ($\gamma$ is the Barbero-Immirzi parameter), that satisfy the Poisson bracket $\left\{c,p\right\}=8\pi\gamma G/3$. From now on, we will set $8\pi G/3 = 1$ and $\gamma = 1$, so that: $$\left\{c,p\right\}=1$$

The Hamiltonian constraint for the FRW model which describes a flat universe in the radiation dominated era is
\begin{equation}
\label{HamiltonianConstraint}
H=-c^{2}\sqrt{p}+\frac{E^{2}}{\sqrt{p}}\approx 0
\end{equation}

The quantity $E$ is related to the magnitude of the electric field, it has been assumed to be sufficiently small so as not to cause significant anisotropy \cite{Bojowald1}, however, it could be argued to be a spacelike average of the electric field's magnitude. As in the interior of a neutral plasma, the electric field does not point somewhere specifically but it certainly is not zero, without breaking isotropy. $E$ is coupled  to the matter content of the universe for the radiation dominated era, in which the energy density  $\rho=E^{2}/a^{4}$  dominates over the other contributions to the total energy density.

\subsection{Electric Time}

One approach to avoid (rather than solve) the problem of time is to identify an internal time as a functional of the canonical variables and then, solve the canonical constraints before the quantization of the system \cite{isham}.

Consider the constraint (\ref{HamiltonianConstraint}) for the system given by the canonically conjugated pairs $c,p$ and $A,E$. The electric field satisfies
\begin{equation}
\label{PoissonAandE}
\left\{A,E\right\}=1
\end{equation}
with $A$ related to the vector potential, and canonically conjugated to the electric field. As the field $A$ doesn't appear in the Hamiltonian constraint, then $E$ is a constant of motion and $A$ could be chosen as an internal time \cite{Bojowald1}.

To determine if $A$ is a good election of internal time, we need to know if that field grows monotonically respect the proper time. We have $\left\{A,H\right\}=2E/\sqrt{p(t)}$, and from this we obtain $A(t)=\sqrt{8Et}$. Therefore, for positive $t$ and $E$, the field $A$ is a reasonable choice of internal time \cite{Tesis}.

Now we have an evolution parameter, the constraint \ref{HamiltonianConstraint} is solved for the electric field, \textit{i.e.} the canonically conjugated moment of $A$. The result is 
$
%\label{Constraint_Solution}
E(c,p)=\pm c\sqrt{p}
$ \cite{Bojowald1}. As we need that $E$ has positive value (and evidently $c$ must be positive), we choose the plus sign 
%in (\ref{Constraint_Solution}) 
such that
\begin{equation}
\label{Classical_E}
E(c,p)=c\sqrt{p}.
\end{equation}
% Since we choose to describe time in terms of the change of $A$,  
On the other hand, a function invariant under the action of the Hamiltonian constraint has $\frac{d\ }{d t}f(c,p,A,E)=0$, thus
\begin{equation}
0=\{f,H\}+\frac{\partial f}{\partial t}=\{f,c\sqrt{p}-E\}\frac{2E}{\sqrt{p}}+\partial_t f.
\end{equation}
If we use it first for $f=A$, we can use the result to write the equations of motion for any function without referring to coordinate time ``$t$"
\begin{equation}
0=\frac{2E}{\sqrt{p}}\Big(\{f(c,p),c\sqrt{p}\}+ \partial_Af(c,p)\Big).
\end{equation}
Here we see that $-E(c,p)$ is the generator of $A$-evolution, and provides classical equations of motion for $c(A)$ and $p(A)$.

The Hamiltonian equations of motion for the classical system are
\begin{equation}
\label{Classical_p}
\partial_A p=\left\{p,-E(c,p)\right\}=\sqrt{p}
\end{equation}
  
\begin{equation}
\label{Classical_c}
\partial_A c=\left\{c,-E(c,p)\right\}=-\frac{c}{2\sqrt{p}}
\end{equation}
with solutions
\begin{equation}
\label{Classical_p_Sol}
p(A)=\frac{1}{4}A^{2}
\end{equation}
and
\begin{equation}
\label{Classical_c_Sol}
c(A)=\frac{2E}{A}.
\end{equation}

From the definition $p=a^{2}$ and the relation between internal time and proper time, we observe that (\ref{Classical_p_Sol}) is an equivalent solution for the cosmic scale factor in the radiation dominated era and, as in conventional cosmological approach, we obtain $a(t)\propto t^{1/2}$.

\section{Effective Equations of the Quantum FRW Flat Universe in the Radiation Dominated Era}

\subsection{Effective Theories}
\label{III.A}

We want to obtain quantum corrections for classical theories. For this, we will follow the procedure of canonical effective equations \cite{PaperTesisAureliano}. In this approach, classical variables are the mean values of fundamental operators $x^{i}=\left\langle\hat{x}^{i}\right\rangle$, which satisfies $\left\{x^{i},x^{j}\right\}=\epsilon^{ij}$.

For quantum variables, \textit{central moments} are introduced, defined as the 
\begin{dmath}
\label{CentralMoments}
G^{i_1\dots i_n}=\left\langle (\hat{x} ^{(i_1} - x^{(i_1})\dots (\hat{x} ^{i_n)} - x^{i_n)})\right\rangle
\end{dmath}
the fully symmetric product of quantum fluctuations. The central moments satisfies $\left\{x^{i},G^{i_{1}\dots i_{n}}\right\}=0$, and appropriate relations (explicitly obtained in \cite{PaperTesisAureliano}\cite{Bojowald5}) between them. As a consequence of its quantum nature, this variables (actually of order $G^{i_1\dots i_n}\propto\hbar^{n/2}$ \cite{PaperTesisAureliano}) must obey uncertainty relations.  

Time evolution is generated by the \textit{Quantum Hamiltonian}, defined as the mean value of the Hamiltonian Operator $$H_{Q}:=\hat{\left\langle H \right\rangle}$$ and expressed in terms of the quantum phase space variables, by Taylor expanding $\hat{\left\langle H \right\rangle}$, as
\begin{equation}
\label{QuantumHamiltonian}
H_Q=\sum_{\substack{n}} \frac{1}{n!}H,_{i_1\dots i_n}G^{i_1\dots i_n}.
\end{equation}
The relation between quantum mechanics from the geometric phase space and the Hilbert space is stated in \cite{schilling} $$\left\{\hat{\left\langle F \right\rangle},\hat{\left\langle K \right\rangle}\right\}=\frac{1}{i\hbar}\left\langle\left[\hat{F},\hat{K}\right]\right\rangle,$$ and it allows us to obtain equations of motion for the variables of the description \cite{PaperTesisAureliano} $$\dot{x}^{i}=\left\{x^{i},H_Q\right\},$$ $$\dot{G}^{i_1\dots i_n}=\left\{G^{i_1\dots i_n},H_Q\right\}.$$ However, infinitely many terms appear in these equations and it becomes necessary some sort of approximation, i.e. to reduce them to a finite number. We call a theory {\it effective} if it accounts for the physics we want to describe in easy terms, in an effective way. In order to achieve an effective description, we perform a semi classical approximation of order $\hbar$. This implies to evaluate (\ref{QuantumHamiltonian}) up to $n=2$. The Quantum Hamiltonian becomes
\begin{equation}
\label{semiclassicalQH}
H_{Q}=H+\frac{1}{2}H_{,ij}G^{ij}
\end{equation}
where $(i,j)$ takes values on phase space coordinates $(q,p)$. Then, the equations of motion are

\begin{equation}
\label{ClassMotEq}
\dot{x}^{i}=\left\{x^{i},H\right\}+\frac{1}{2}\left\{x^{i},H_{,jk}\right\}G^{jk}
\end{equation}

\begin{equation}
\label{QuantMotEq}
\dot{G}^{ij}=\frac{1}{2}H_{,kl}\left\{G^{ij},G^{kl}\right\}
\end{equation}

As we can see in (\ref{ClassMotEq}), the classical equation has an additional term of $\hbar$ order. 
Employing the commutation relation between the central moments 
$$\left\{G^{ij},G^{kl}\right\}=\epsilon^{ik}G^{jl}+\epsilon^{il}G^{jk}+\epsilon^{jl}G^{ik}+\epsilon^{jk}G^{il}$$ 
and the definition of symplectic vector fields (also known as \textit{Hamiltonian} vector fields \cite{Arnold}) in terms of coordinates 
$$X_{F}=\left\{x^{k},F\right\}\partial_{k}=\epsilon^{kl}(\partial_{l}F)\partial_{k}.$$ 
Assuming the dynamics of the central moments can be described in terms of the evolution of the {\it classical} phase space variables $\dot G^{ij}\rightarrow\dot x^k\partial_k G^{ij}$, equation (\ref{QuantMotEq}) becomes 
$$X^{k}_{H}\partial_{k}G^{ij}-X^{i}_{H},_{l}G^{jl}-X^{j}_{H},_{l}G^{il}=0$$ 
which is the Lie derivative of $G^{ij}$ in the direction of the symplectic vector field $X_{H}$. 
Then, we obtain the quantum corrections by solving
\begin{equation}
\label{G_LieDerivative}
\pounds_{X_{H}}G^{ij}=0.
\end{equation}

We propose solutions with the structure
\begin{equation}
\label{Proposed_G}
G^{ij}=\sum_{\substack{A,B}} X^{i}_{F^{A}} X^{j}_{F^{B}} G_{AB}
\end{equation}
where $(A,B)$ run for $2N$ functions $F^A$, a new set of coordinates of the phase space. $G_{AB}$ are functions of the constants of motion and $X^{i}_{F^{A}}$ are symplectic vectors associated to functions $F^{A}$. These functions must satisfy 
\begin{equation}
\label{Function_Condition}
\left\{F^{A},H\right\}=
{ constants\ of\ motion}.  \end{equation}
Then, by getting the $2N$ functions $F^{A}$ and computing its associated symplectic vector fields, we can construct the quantum correction terms. To obtain the quantities $G_{AB}$, we have to impose some conditions on the equations of motion. 

As we mentioned previously, quantum variables must obey uncertainty relations. At order $\hbar$ this relation becomes 
\begin{equation}
\label{SC_Uncertainty}
G^{qq}G^{pp}-(G^{qp})^{2}\geq\frac{\hbar^{2}}{4} \end{equation}

We also require the equivalence of the equations of motion in the effective subspace and the full setting.  Then, if we defined the \textit{effective Hamiltonian} $$H_{eff}=\left.H_{Q}\right|_{G=G\left(x^{i}\right)}$$ the condition to impose is
\begin{equation}
\label{Q_equal_E}
\left. \left\{x^{i},H_{Q}\right\}\right|_{G=G\left(x^{i}\right)}=\left\{x^{i},H_{eff}\right\}_{eff}
\end{equation}
with the corrected \cite{TesisAureliano} Poisson Bracket given by
\begin{equation}
\label{Corrected_PB}
\left\{x^{i},x^{j}\right\}_{eff}=\epsilon^{ij}(1+\kappa)
\end{equation}
Here, $\kappa$ is a function of the phase space variables of order $\hbar$ and we determine its value using (\ref{Corrected_PB}) in equation (\ref{Q_equal_E}). 
Since the Poisson Bracket of three functions satisfies the Jacobi identity \cite{Arnold}, we have in the effective subspace
\begin{equation}
\label{Jacobi}
\left\{x^{i},\dot{x}^{j}\right\}_{eff}-\left\{x^{j},\dot{x}^{i}\right\}_{eff}+\epsilon^{ij}\left\{H_{eff},\kappa\right\}_{eff}=0.
\end{equation}

\subsection{The Quantum Cosmological Model}

In the following section we are going to use the techniques developed in \ref{III.A} on the cosmological model of section \ref{II}.

We choose to work with canonical variables $\left\langle\hat{p}\right\rangle\rightarrow p(A)$ and $\left\langle\hat{c}\right\rangle\rightarrow c(A)$. The quantum Hamiltonian is
\begin{equation}
\label{Quantum_E}
-E_{Q}=-c\sqrt{p}-\frac{1}{2\sqrt{p}}G^{cp}+\frac{1}{8}\frac{c}{p^{3/2}}G^{pp}+\mathcal{O}(\hbar^{2})
\end{equation}
where evolution in $A$ is generated by $-E_{Q}$. It provides the corrected equations of motion
\begin{equation}
\label{Quantum_p}
\partial_A p=\left\{p,-E_{Q}\right\}=\sqrt{p}-\frac{1}{8p^{3/2}}G^{pp}
\end{equation}
  
\begin{equation}
\label{Quantum_c}
\partial_A c=\left\{c,-E_{Q}\right\}=-\frac{c}{2\sqrt{p}}+\frac{1}{4p^{3/2}}G^{cp}-\frac{3}{16}\frac{c}{p^{5/2}}G^{pp}.
\end{equation}
In order to determine the effective equations of the system we ought to solve for $G^{cc}$, $G^{cp}$ and $G^{pp}$ using (\ref{Proposed_G}).

The first step is to find $2$ functions $F^{A}$ that satisfy (\ref{Function_Condition}). 
The Hamiltonian itself is a good choice because $\left\{E,E\right\}=0$. 
The second function must obey $\left\{F,E\right\}=1$. 
This results in the differential equation 
$$\left(\frac{c}{2\sqrt{p}}\partial_{c}-\sqrt{p}\partial_{p}\right)F=1$$ 
with solution
\begin{equation}
\label{F_Function}
F(c,p)=-2\sqrt{p}
\end{equation}

Now, we compute the symplectic vector fields associated to $E$ and $F$
\begin{equation}
\label{Symplectic_E}
X_{E}=(\partial_{p}E)\partial_{c}-(\partial_{c}E)\partial_{p}=\frac{c}{2\sqrt{p}}\partial_{c}-\sqrt{p}\partial_{p}
\end{equation}

\begin{equation}
\label{Symplectic_F}
X_{F}=(\partial_{p}F)\partial_{c}-(\partial_{c}F)\partial_{p}=-\frac{1}{\sqrt{p}}\partial_{c}
\end{equation}
and inserting (\ref{Symplectic_E},\ref{Symplectic_F}) in (\ref{Proposed_G}) we obtain the correction terms
\begin{equation}
\label{Gcc}
G^{cc}=G_{{}_{EE}}\frac{c^{2}}{4p}-G_{{}_{EF}}\frac{c}{p}+G_{{}_{FF}}\frac{1}{p}
\end{equation}

\begin{equation}
\label{Gcp}
G^{cp}=-G_{{}_{EE}}\frac{c}{2}+G_{{}_{EF}}
\end{equation}

\begin{equation}
\label{Gpp}
G^{pp}=G_{{}_{EE}}p
\end{equation}

From the restrictions to the  effective dynamics we obtain 
$$\kappa(c,p)=\frac{3E}{8p}G_{{}_{EE}}'+\frac{1}{4p}G_{{}_{EE}}+\frac{1}{2\sqrt{p}}G_{{}_{EF}}'$$ 
where prime denotes derivation respect to $E$. 
Additionally, the Jacobi identity (\ref{Jacobi}) provides \begin{equation}
\label{Diff_Eq_E}
(4p^{3/2}+1)\left(\frac{3E}{2}G_{{}_{EE}}'+G_{{}_{EE}}+\sqrt{p}G_{{}_{EF}}'\right)=0
\end{equation}
As the first bracket term can't be zero, we obtain an ordinary differential equation for $G_{{}_{EE}}$ and $G_{{}_{EF}}$ with solutions
\begin{equation}
\label{Gee}
G_{{}_{EE}}=\alpha E^{-2/3}
\end{equation}
\begin{equation}
\label{Gef}
G_{{}_{EF}}=\beta
\end{equation}
where $\alpha$ and $\beta$ are integration constants to be determined by means of the uncertainty relation. 
Evaluation of the totally saturated uncertainty relation (\ref{SC_Uncertainty}) in the effective subspace gives 
$$G_{{}_{EE}}G_{{}_{FF}}-\left(G_{{}_{EF}}\right)^{2}=\frac{\hbar^{2}}{4}$$ 
where we can observe that functions $G_{{}_{EE}}$ and $G_{{}_{FF}}$ can't vanish. 
In terms of the integration constants we have $$\alpha E^{-2/3}G_{{}_{FF}}-\beta^{2}=\frac{\hbar^{2}}{4}$$ a relation which doesn't provide us the value of $\alpha$ nor $\beta$. 
We can only say that this integration constants are of order $\hbar$ and that $\alpha$ must be greater than zero.

We can finally express the quantum correction terms (\ref{Gcc}, \ref{Gcp}, \ref{Gpp}) in function of the  integration constants $\alpha$ and $\beta$. These result in
\begin{equation}
\label{Effective_Gcc}
G^{cc}=\frac{E^{-2/3}\alpha c^{2}}{4p}-\beta\frac{c}{p}+\frac{E^{2/3}}{\alpha p}\left(\frac{\hbar^{2}}{4}+\beta^{2}\right)
\end{equation}
\begin{equation}
\label{Effective_Gcp}
G^{cp}=-\frac{E^{-2/3}\alpha c}{2}+\beta
\end{equation}
\begin{equation}
\label{Effective_Gpp}
G^{pp}=E^{-2/3}\alpha p
\end{equation}

\section{Solutions}

\subsection{The Big Bouncing Scenario}

Equation (\ref{Quantum_p}) is related to the evolution of the cosmic scale factor. We can write it as
\begin{equation}
\label{Motion_p}
\frac{\partial p}{\partial A}=\sqrt{p}-\frac{E^{-2/3}}{8\sqrt{p}}\alpha
\end{equation}

Now, we propose a power series in $\alpha(\hbar)$ as the solution. 
That is $$p(A)=\sum\limits_{n}p_{{}_{n}}(A)\alpha^{n}$$ 
with coefficients $p_{{}_{n}}(A)=(n!)^{-1}\left.\partial_{\alpha}^{n}p(A)\right|_{\alpha=0}$. 
The zeroth order term
\begin{eqnarray}
\label{Order_zero}
\nonumber \frac{\partial p_{{}_{0}}}{\partial A}&=&\sqrt{p_{{}_{0}}}\\
p_{{}_{0}}&=&\frac{A^{2}}{4}
\end{eqnarray}
agree with the classical solution of the model. 
The term with $n=1$ is
\begin{eqnarray}
\label{Order_one}
\nonumber \frac{\partial p_{{}_{1}}}{\partial A}&=&\frac{p_{{}_{1}}}{2\sqrt{p_{{}_{0}}}}-\frac{E^{-2/3}}{8\sqrt{p_{{}_{0}}}}\\
\nonumber {}&=&\frac{p_{{}_{1}}}{A}-\frac{E^{-2/3}}{4A}\\
p_{{}_{1}}&=&A\lambda+\frac{-E^{2/3}}{4}
\end{eqnarray}
with $\lambda$ and integration constant. Then, the solution for $p(A)$ at order $\hbar$ is 
\begin{equation}
\label{P_solution}
p(A)=\left(\frac{A}{2}+\lambda\alpha\right)^{2}+\frac{E^{-2/3}}{4}\alpha.
\end{equation}
Clearly, in the limit $\alpha(\hbar)\rightarrow 0$, (\ref{P_solution}) reduces to the classical solution. As we can see, when the universe becomes smaller, the minimum size it takes is $$p_{\min}=\frac{E^{-2/3}}{4}\alpha,$$  
this implies
%, as we can see in picture (\ref{Plot}) 
the elimination of the initial singularity appearing in FRW's classical models and its substitution by a term of quantum nature.

%\begin{figure}[hbtp!]  \centering  \includegraphics[height=7cm]{Graph1} \caption{Evolution of $p(A)$} \label{Plot}  \end{figure}

\subsection{The Effective Action}

Following our method, effective equations of motion are obtained evaluating equations (\ref{Quantum_p}, \ref{Quantum_c}) in the effective subspace. Quantum correction terms are given by equations (\ref{Effective_Gcc}, \ref{Effective_Gcp} and \ref{Effective_Gpp}). 

Alternatively we can derive the effective equations of motion starting directly from the effective Hamiltonian, \textit{i.e.} the quantum Hamiltonian (\ref{Quantum_E}) evaluated on the effective subspace. Then, we have

\begin{equation}
\label{Effective_E}
E_{eff}=c\sqrt{p}-\frac{1}{2\sqrt{p}}\left(\frac{3}{4}c^{1/3}p^{-1/3}\alpha-\beta\right)
\end{equation}

From this equation, we can write the corrected Hamiltonian constraint
\begin{equation}
\label{Effective_H_Constraint}
H_{eff}=-c^{2}\sqrt{p}+\frac{E^{2}}{\sqrt{p}}-\frac{c}{\sqrt{p}}\left(\frac{3}{4}c^{1/3}p^{-1/3}\alpha-\beta\right)
\end{equation}

If we take the classical limit $\hbar\rightarrow 0$ in equation (\ref{Effective_H_Constraint}), equation (\ref{HamiltonianConstraint}) could be recovered. Now, from the relation between Ashtekar's variables and the time dependent cosmic scale factor, we compute $\left\{p,H_{eff}\right\}$, and from this, we obtain the corrected $c(t)$ in terms of metric variables

\begin{equation}
\label{Corrected_c_metric}
c=\dot{a}+\frac{1}{2a^{2}}\left(\left(\frac{\dot{a}}{a^{2}}\right)^{1/3}\alpha-\beta\right)
\end{equation}

With a Legendre transformation we compute the effective Lagrangian
\begin{equation}
\label{Eff_Lagrangian}
\mathcal{L}_{eff}=-\frac{a^{3}R}{6}+\frac{E^{2}}{a}
-\frac{7}{4}a^{-1/3}{\mathcal H}^{4/3}\alpha\end{equation}
where $\mathcal{H}=\frac{\dot a}{a}$ is the Hubble parameter. 
This expression can be rewritten in terms of $R$, the Ricci scalar, and the Kretschmann scalar for the spatially flat FRW metric. 
Since the Riemann can be expressed in terms of projectors, we can compute easily the Kretschmann
\begin{dmath}
R_{\mu\nu}{}^{\lambda\kappa}=-2\frac{\ddot{a}}{a} \frac{(\delta^t_\mu\delta^i_\nu-\delta^i_\mu\delta^t_\nu)(\delta_t^\lambda\delta_i^\kappa-\delta_i^\lambda\delta_t^\kappa)}{2}
-2\Big(\frac{\dot{a}}{a}\Big)^2 \frac{(\delta^i_\mu\delta^j_\nu-\delta^j_\mu\delta^i_\nu)(\delta_i^\lambda\delta_j^\kappa-\delta_j^\lambda\delta_i^\kappa)}{4},
\end{dmath} with this, we can use $$\Big(\frac{\dot{a}}{a}\Big)^2=-\frac{R}{12}\pm\sqrt{\frac{K}{24}-\frac{R^2}{144}}$$
where $R=R_{\mu\nu}{}^{\mu\nu}$ and $K=R_{\mu\nu}{}^{\lambda\kappa}R_{\lambda\kappa}{}^{\mu\nu}.$ The proposed effective Lagrangian for GR is
\begin{equation}
\label{FullEff_Lagrangian}
\mathcal{L}_{eff}=\frac{\sqrt{g}R}{6}-\frac{\langle \vec E^{2}\rangle}{\sqrt{g}}-\frac{7}{4}\sqrt{g}^{-\frac{1}{9}}\alpha\left(-\frac{R}{12}\pm\sqrt{\frac{K}{24}-\frac{R^2}{144}}\right)^{\frac{2}{3}}
\end{equation}
but correction terms do not appear as generally covariant but with an abnormal weight $w=-\frac{1}{9}$.

\section{Conclusions}

We compute effective equations for a cosmological model where radiation is the only content in order to study significant quantum effects in the early universe. 

We develop a method to compute effective equations for quantum systems. It exploits the symmetries of the equations of motion for the central moments under the flux generated by the symplectic field associated to the Hamiltonian.

The problem of time is avoided by choosing the field canonically conjugated to the radiation source field as time parameter, then, evolution is provided by the electric field's conjugate variable.

In the context of Loop Quantum Cosmology it has become clear that rather than a Big Bang, LQC predicts a Big Bounce. Here, we do not regularize the cosmological variables, yet, instead of a Big Bang, we get a family of bouncing solutions for the cosmic scale factor which drop out the initial singularity of the classical model. We prove that there is no big bang solution, but cannot yet affirm that this will happen generically when studying effective equations with other physical time choices. 

We also obtain the effective Lagrangian of the model as the classical Lagrangian with additional quantum correction terms.
We can see that the correction term doesn't have the correct weight. This appears to be a generic result coming from the quantization procedure, it could be said that covariance is broken at Planck regimes, more specifically diffeomorphisms that change the volume form appear to be broken. However, this is not unexpected, because the quantization is performed over a symmetric background. The background breaks the symmetry.

We are curious about whether the effective action can be improved by imposing some conditions on the equations of motion for metric perturbations around the homogeneous background.
Anyway, further analysis of this kind is  possible in wider contexts.

\end{document}